\documentclass{appolb}
\usepackage{epsfig}
\title{DGLAP EVOLUTION OF TRUNCATED MOMENTS OF PARTON DENSITIES WITHIN TWO 
DIFFERENT APPROACHES.}

\author{Dorota Kotlorz $^1$, Andrzej Kotlorz $^2$
\address{$^1$Department of Physics Ozimska 75, $^2$Department of 
Mathematics Luboszycka 3, Technical University of Opole, 
45-370 Opole, Poland, e-mail $^1$: {\tt dstrozik@po.opole.pl}}}

\begin{document}
\pagestyle{plain}
\eqsec
\maketitle

\begin{abstract}
We solve the LO DGLAP QCD evolution equation for truncated Mellin moments of 
the nucleon nonsinglet structure function. The results are compared with those, 
obtained in the Chebyshev-polynomial approach for $x$-space solutions.
Computations are performed for a wide range of the truncation point 
$10^{-5}\leq x_0\leq 0.9$ and $1\leq Q^2\leq 100 {\rm GeV}^2$. The agreement
is perfect for higher moments ($n\geq 2$) and not too large $x_0$ ($x_0\leq
0.1$), even for a small number of terms in the truncated series ($M=4$). The 
accuracy of the truncated moments method increases for larger $M$ and
decreases very slowly with increasing $Q^2$. For $M=30$ the relative error in a
case of the first moment at $x_0\leq 0.1$ and $Q^2=10 {\rm GeV}^2$ doesn't
exceed 5\% independently on the shape of the input parametrisation. This is
a quite satisfactory result. Using the truncated moments approach one can avoid 
uncertainties from the unmeasurable $x\rightarrow 0$ region and also study 
scaling violations without making any assumption on the shape of input 
parametrisation of parton distributions. Therefore the method of truncated 
moments seems to be a useful tool in further QCD analyses.

\end{abstract}


\section{Introduction}

The DGLAP evolution \cite{b1} is the most familiar resummation
technique, which describes scaling violations of parton densities.
Measurements of deep-inelastic scattering structure functions of the nucleon
allow the determination of free parameters of the input parton distributions 
and the verification of so called sum rules. There exist different sum rules 
for unpolarised and polarised structure functions which refer to the moments 
of the structure functions. From a phenomenological point of view however, QCD
tests based on moments $\int_{0}^{1} dx x^{n-1} F(x,Q^2)$ is
unreliable. The limit $x\rightarrow 0$, which implies that the invariant
energy $W^2$ of the inelastic lepton-hadron scattering becomes infinite
($W^2=Q^2(1/x-1)$) will never be attained experimentally. In the theoretical 
approach to structure functions there are two ways to avoid the problem of 
dealing with the unphysical region $x\rightarrow 0$. The first one is to work in
$x$-space and obtain directly the evolution of parton distributions (not of
their moments). Then one has integro-differential equations (e.g. DGLAP one)
in $x$ and $Q^2$ but the integration over $x$ goes for $x\geq x_0$. In this
case an extrapolation to the unmeasurable $x\rightarrow 0$ region is unneeded. 
The second way is using evolution equations for truncated moments of structure
functions $\int_{x_1}^{x_2} dx x^{n-1} F(x,Q^2)$ instead of for full
moments. In the usually used method of solving QCD evolution equations, one 
takes the Mellin (full) transform of these equations and obtains analytical 
solutions. Then after the inverse Mellin transform (performed numerically) one
has suitable solutions of the original equations in $x$-space. In this way e.g. 
in a case of DGLAP approximation, the differentio-integral equations for parton 
distributions $q(x,Q^2)$ change after the Mellin transform into simple differential 
and diagonalised ones in the moment space $n$. The only problem is knowledge of 
the input parametrisation for the whole region $0\leq x \leq 1$ what is necessary 
in the determination of the initial moments of the distribution functions. 
Using truncated moments approach one can avoid uncertainties from the 
unmeasurable $x\rightarrow 0$ region and also obtain important theoretical 
results incorporating perturbative QCD effects at small $x$, which could be 
verified experimentally. Truncated moments of parton distributions in solving 
DGLAP equations have been presented in \cite{b5}. Authors have shown that the 
evolution equations for truncated moments though not diagonal can be solved 
with a quite good precision for $n\geq 2$. This is because each $n$-th truncated 
moment couples only with ($n+j$)-th ($j\geq 0$) truncated moments. In \cite{b6} 
the truncated moments method has been adopted to double logarithmic $ln^2x$ 
resummation. There is a number of papers in which the most known methods for 
solving the $Q^2$ evolution equations for parton distributions have been 
reviewed (see e.g. \cite{b7},\cite{b8}). Authors compare the DGLAP framework 
for the full Mellin moments method with brute-force or Laguerre-polynomial 
approaches, used for $x$-space version of the evolution equation. In this paper 
we compare the solutions of LO DGLAP $Q^2$ evolution equations written for the 
truncated Mellin moments of the structure functions with those, obtained by
using the Chebyshev-polynomial method in the $x$-space. In both these approaches we 
compute the truncated moments $\int_{x_0}^{1} dx x^{n-1} F(x,Q^2)$. As a test 
structure function $F(x,Q^2)$ we take two different spin-like nonsinglet parton
distributions. We perform the computations for a wide range of the truncation 
point $10^{-5}\leq x_0\leq 0.9$ and $1\leq Q^2\leq 100 {\rm GeV}^2$. In the
next section we briefly recall an idea of the evolution equation for truncated 
moments of parton distributions. The main topic of our paper i.e. the 
comparison of the Chebyshev-polynomial and truncated moments techniques in 
solving the LO DGLAP evolution equation for the nonsinglet structure function 
is presented in Section 3. Finally, Section 4 contains conclusions.

\section{Truncated Mellin moments of the nonsinglet structure function 
$q^{NS}(x,t)$ within LO DGLAP approach.}

For (full) Mellin moments of parton distributions $f(x,Q^2)$
\begin{equation}\label{r3.1}
\bar{f}(n,Q^2)=\int\limits_0^1 dx x^{n-1} f(x,Q^2)
\end{equation}
the DGLAP evolution equation can be solved analytically. This is because one 
obtains in the moment space $n$ simple diagonalised differential equations. The 
only problem is the knowledge of the input parametrisation for the whole region 
$0\leq x\leq 1$, what is necessary in the determination of the initial moments 
$\bar{f}(n,Q^2=Q_0^2)$:
\begin{equation}\label{r3.2}
\bar{f}(n,Q_0^2)=\int\limits_0^1 dx x^{n-1} f(x,Q_0^2).
\end{equation}
Using the truncated moments approach one can avoid the uncertainties from the 
region $x\rightarrow 0$, which will never be attained experimentally. The 
derivation of the DGLAP equations for truncated moments of parton distributions 
has been presented in \cite{b5}. The evolution equations for truncated moments
$\bar{f}(x_0,n,Q^2)$
are not diagonal and therefore solving this problem is not so easy like in a
case of the full-moments technique. Nevertheless this method has an
advantage over other approaches, based not only on the cut-off for unphysical
region $x\rightarrow 0$. The technique of truncated moments within DGLAP
approximation enables namely to study scaling violations without making any
assumption on the shape of the input parametrisation of parton distributions.
While the solution of the evolution equations in the $x$-space requires
knowledge of inputs $f(x,Q_0^2)$ with many parameters (fitted in detailed
comparison with the data), the initial values of truncated moments can be
obtained directly by data. Following the authors of \cite{b5}, we have found
the LO DGLAP evolution equation for the truncated at $x_0$ Mellin moment of
the nonsinglet structure function $q^{NS}(x,Q^2)$ in a form:
\begin{equation}\label{r3.4}
\frac{d\bar{q}^{NS}(x_0,n,t)}{dt}=\frac{\alpha_s(t)}{2\pi}
\int\limits_{x_0}^1 dy y^{n-1} q^{NS}(y,t) G_{n}\left(\frac{x_0}{y}\right).
\end{equation}
$\bar{q}^{NS}(x_0,n,t)$ is the truncated at $x_0$ moment of the nonsinglet
structure function:
\begin{equation}\label{r3.5}
\bar{q}^{NS}(x_0,n,t)=\int\limits_{x_0}^1 dx x^{n-1} q^{NS}(x,t),
\end{equation}
where
\begin{equation}\label{r2.4}
t\equiv \ln\frac{Q^2}{\Lambda_{QCD}^2}
\end{equation}
and 
\begin{equation}\label{r3.6}
G_n\left(\frac{x_0}{y}\right)\equiv\int\limits_{x_0/y}^1 dz z^{n-1}
P_{qq}(z).
\end{equation}
For $x_0=0$ the kernel $G_n(x_0/y)$ is simply equal to the anomalous
dimension $\gamma_{qq}(n)$:
\begin{equation}\label{r2.5}
\gamma_{qq}(n)=\int\limits_0^1 z^{n-1} P_{qq}(z) dz.
\end{equation}
Expanding the $G_n$ in Taylor series around $y=1$, one has
\begin{eqnarray}\label{r3.7}
G_n\left(\frac{x_0}{y}\right) = \gamma_{qq}(n)-\frac{4}{3}\sum\limits_{k=0}^{\infty} 
[2\sum\limits_{i=n+2}^{\infty}\frac{(i+k-1)!}{i!}x_0^i \nonumber\\
 + \frac{(n+k-1)!}{n!}
(x_0^n+\frac{n+k}{n+1} x_0^{n+1})]\sum\limits_{p=0}^{k}\frac{(-1)^p
y^p}{p!(k-p)!}.
\end{eqnarray}
Truncating the above expansion at order $M$ and using the following relation
\begin{equation}\label{r3.8}
\sum\limits_{k=0}^{M}\sum\limits_{p=0}^{k} \longrightarrow 
\sum\limits_{p=0}^{M}\sum\limits_{k=p}^{M}
\end{equation}
one can find that the evolution equation (\ref{r3.4}) becomes
\begin{equation}\label{r3.9}
\frac{d\bar{q}^{NS}(x_0,n,t)}{dt}=\frac{\alpha_s(t)}{2\pi}
\sum\limits_{p=0}^M C_{pn}^{(M)}(x_0) \:\bar{q}^{NS}(x_0,n+p,t)
\end{equation}
and
\begin{equation}\label{r3.9a}
G_n\left(\frac{x_0}{y}\right) = \sum\limits_{p=0}^M C_{pn}^{(M)}(x_0) \:
y^p,
\end{equation}
where
\begin{eqnarray}\label{r3.10}
C_{pn}^{(M)}(x_0) = \gamma_{qq}(n)\delta_{p0} - \frac{4}{3}\sum\limits_{k=p}^{M}
\frac{(-1)^p}{p!(k-p)!}
\:[\:2\sum\limits_{i=n+2}^{\infty}\frac{(i+k-1)!}{i!}x_0^i\nonumber\\
 + \frac{(n+k-1)!}{n!}(x_0^n+\frac{n+k}{n+1} x_0^{n+1})\:].
\end{eqnarray}
Note that the evolution equations for truncated moments (\ref{r3.9}),
(\ref{r3.10}) are not diagonal but each $n$-th moment couples only with 
($n+p$)-th ($p\geq 0$) moments. As it was shown in \cite{b5} the series of 
couplings to higher moments is convergent and furthermore the value of 
($n+p$)-th moments decreases rapidly in comparison to the $n$-th moment. Hence 
one can retain from (\ref{r3.9}) the closed system of $M+1$ equations:
\begin{displaymath}
\frac{d\bar{q}^{NS}(x_0,N_0,t)}{dt}=\frac{\alpha_s(t)}{2\pi}
[C_{0,N_0}^{(M)}(x_0) \bar{q}^{NS}(x_0,N_0,t)
\end{displaymath}
\begin{displaymath}
+ C_{1,N_0}^{(M)}(x_0) \bar{q}^{NS}(x_0,N_0+1,t) + \\
...+ C_{M,N_0}^{(M)}(x_0) \bar{q}^{NS}(x_0,N_0+M,t)]
\end{displaymath}
\begin{displaymath}
\frac{d\bar{q}^{NS}(x_0,N_0+1,t)}{dt}=\frac{\alpha_s(t)}{2\pi}
[C_{0,N_0+1}^{(M-1)}(x_0) \bar{q}^{NS}(x_0,N_0+1,t)
\end{displaymath}
\begin{displaymath}
+ C_{1,N_0+1}^{(M-1)}(x_0) \bar{q}^{NS}(x_0,N_0+2,t) +
...+ C_{M-1,N_0+1}^{(M-1)}(x_0) \bar{q}^{NS}(x_0,N_0+M,t)]
\end{displaymath}
\begin{displaymath}
...
\end{displaymath}
\begin{equation}\label{r3.11}
\frac{d\bar{q}^{NS}(x_0,N_0+M,t)}{dt}=\frac{\alpha_s(t)}{2\pi}
C_{0,N_0+M}^{(0)}(x_0) \bar{q}^{NS}(x_0,N_0+M,t).
\end{equation}
$N_0$ denotes the lowest moment in calculations. The above system can be
solved numerically like a standard coupled differential equations using 
the Runge-Kutta method. We have also found an analytical solution of (\ref{r3.11}) 
in the form:
\begin{eqnarray}\label{r3.12}
\bar{q}^{NS}(x_0,i,t) = \left(\bar{q}^{NS}(x_0,i,t_0)
 - \sum\limits_{k=i+1}^{N_0+M}
A_{ik}(x_0)\bar{q}^{NS}(x_0,k,t_0)\right)\nonumber\\
\times \exp\left(\frac{\alpha_s}{2\pi}D_{ii}^{(M)}(x_0)
(t-t_0)\right) + \sum\limits_{k=i+1}^{N_0+M} A_{ik}(x_0)\bar{q}^{NS}(x_0,k,t)
\end{eqnarray}
for $\alpha_s=$const and
\begin{eqnarray}\label{r3.13}
\bar{q}^{NS}(x_0,i,t) = \left(\bar{q}^{NS}(x_0,i,t_0) - \sum\limits_{k=i+1}^{N_0+M}
A_{ik}(x_0)\bar{q}^{NS}(x_0,k,t_0)\right)\nonumber\\
\times \exp\left(c_f D_{ii}^{(M)}(x_0)\ln{\frac{t}{t_0}}\right)
+ \sum\limits_{k=i+1}^{N_0+M} A_{ik}(x_0)\bar{q}^{NS}(x_0,k,t)
\end{eqnarray}
for the running $\alpha_s$. Matrix elements $D_{ij}^{(M)}(x_0)$ and
$A_{ij}(x_0)$ are given in Appendix B. For details about properties of
triangular matrices like $D$ see also \cite{b5}. We have made sure that the 
results (\ref{r3.12})-(\ref{r3.13}) agree with the solutions obtained with
the help of the Runge-Kutta method. In the forthcoming chapter we compare predictions for the 
truncated moments $\bar{q}^{NS}(x_0,n,t)$ obtained by solving eq.(\ref{r3.11}) 
with those, computed in the Chebyshev polynomial approach.

\section{Results for truncated moments of the nonsinglet structure function 
$\bar{q}^{NS}(x_0,n,t)$ within LO approximation of the DGLAP approach.}

We solve the system of evolution equations for truncated moments 
(\ref{r3.11}) and compare the results with predictions, obtained in the 
Chebyshev polynomial approach.
The Chebyshev polynomials technique \cite{b15} was successfully used by 
J.Kwieci\'nski in many QCD treatments e.g. \cite{b4},\cite{b12}. Using this 
method one obtains the system of linear differential equations instead of the 
original integro-differential ones. The Chebyshev expansion provides a robust 
method of discretising a continuous problem. This allows computing the parton
distributions for "not too singular" input parametrisation in the whole
$x\in (0;1)$ region. More detailed description of the Chebyshev polynomials
method in the solving the QCD evolution equations is given in Appendix A.
In this paper we use two spin-like input parametrisations of the parton 
distribution $q^{NS}(x,Q_0^2)$ at $Q_0^2=1 {\rm GeV}^2$, namely:
\begin{eqnarray}\label{r4.1}
q^{NS}(x,Q_0^2) &=& a_1 (1-x)^3, \\ \label{r4.2}
q^{NS}(x,Q_0^2) &=& a_2 x^{-0.4}(1-x)^{2.5},
\end{eqnarray}
where constants $a_1$ and $a_2$ are determined by the appropriate sum rules.
More singular at small-$x$ input (\ref{r4.2}) incorporates the latest
knowledge about the low-$x$ behaviour of the polarised structure functions 
\cite{b14}. We start our analysis with a simple test, where the truncation
point $x_0=0$. Then the results should be of course equal to the analytical
ones:
\begin{equation}\label{r4.2a}
\bar{q}^{NS}(n,Q^2)=\bar{q}^{NS}(n,Q_0^2)\left(\frac{\alpha_s(Q_0^2)}
{\alpha_s(Q^2)}\right)^{c_f\gamma_{qq}(n)}.
\end{equation}
$\alpha_s(Q^2)$ is the running coupling and $c_f$ depends on the number of the 
quark flavours $N_f$:
\begin{equation}\label{r2.9}
c_f=\frac{2}{11-\frac{2}{3}N_f}.
\end{equation}
\begin{table}[ht]
\begin{center}
\begin{tabular}{|c|c|c|c|c|}
\hline\hline
$q^{NS}(x,Q_0^2)$ & $Q^2$ & $n$ & $\bar{q}^{NS}(n,Q^2)$ &
$\Delta_{Cheb}\%$ \\ \hline\hline
         &                   & 1 & $2.112\cdot 10^{-1}$ & $< 4\cdot 10^{-1}$ \\ 
\cline{3-5}
         &                   & 2 & $2.820\cdot 10^{-2}$ & $< 4\cdot 10^{-2}$ \\ 
\cline{3-5}
         &        100        & 3 & $7.492\cdot 10^{-3}$ & $< 2\cdot 10^{-1}$ \\ 
\cline{3-5}
         &                   & 4 & $2.732\cdot 10^{-3}$ & $< 3\cdot 10^{-1}$ \\ 
\cline{3-5}
$a_1 (1-x)^3$ &              & 5 & $1.204\cdot 10^{-3}$ & $< 7\cdot 10^{-1}$ \\ 
\cline{2-5}
         &                   & 1 & $2.112\cdot 10^{-1}$ & $< 2\cdot 10^{-1}$ \\ 
\cline{3-5}
         &                   & 2 & $3.296\cdot 10^{-2}$ & $< 2\cdot 10^{-2}$ \\ 
\cline{3-5}
         &        10         & 3 & $9.556\cdot 10^{-3}$ & $< 5\cdot 10^{-2}$ \\ 
\cline{3-5}
         &                   & 4 & $3.709\cdot 10^{-3}$ & $< 2\cdot 10^{-1}$ \\ 
\cline{3-5}
         &                   & 5 & $1.716\cdot 10^{-3}$ & $< 3\cdot 10^{-1}$ \\ 
\hline
         &                   & 1 & $2.112\cdot 10^{-1}$ & $< 2$ \\ 
\cline{3-5}
         &                   & 2 & $2.098\cdot 10^{-2}$ & $< 5\cdot 10^{-2}$ \\ 
\cline{3-5}
         &        100        & 3 & $5.245\cdot 10^{-3}$ & $< 2\cdot 10^{-1}$ \\ 
\cline{3-5}
         &                   & 4 & $1.902\cdot 10^{-3}$ & $< 3\cdot 10^{-1}$ \\ 
\cline{3-5}
$a_2 x^{-0.4}(1-x)^{2.5}$ &  & 5 & $8.502\cdot 10^{-4}$ & $< 2$ \\ 
\cline{2-5}
         &                   & 1 & $2.112\cdot 10^{-1}$ & $< 9\cdot 10^{-1}$ \\ 
\cline{3-5}
         &                   & 2 & $2.452\cdot 10^{-2}$ & $< 2\cdot 10^{-2}$ \\ 
\cline{3-5}
         &        10         & 3 & $6.691\cdot 10^{-3}$ & $< 4\cdot 10^{-2}$ \\ 
\cline{3-5}
         &                   & 4 & $2.583\cdot 10^{-3}$ & $< 9\cdot 10^{-2}$ \\ 
\cline{3-5}
         &                   & 5 & $1.212\cdot 10^{-3}$ & $< 2\cdot 10^{-1}$ \\ 
\hline
\end{tabular}
\caption{Test of the Chebyshev polynomial method: comparison  with analytical 
results of  n-th (full) moments $\bar{q}^{NS}(n,Q^2)$ for different $Q^2$
and input functions $q^{NS}(x,Q_0^2)$.}
\end{center}
\end{table}
Table 1 shows the analytical values of full moments $\bar{q}^{NS}(n,t)$ for
two values of $Q^2$: 10 ${\rm GeV}^2$ and 100 ${\rm GeV}^2$ together with the 
percentage errors for the Chebyshev results $\Delta_{Cheb}\%$:
\begin{equation}\label{r4.3}
\Delta_{Cheb}\% =
\frac{\mid\bar{q}^{NS}(n,t)(analytical)-\bar{q}^{NS}(n,t)(Chebyshev)\mid}
{\bar{q}^{NS}(n,t)(analytical)}\cdot 100\%.
\end{equation}
Note a good agreement of the Chebyshev solutions for 
$\bar{q}^{NS}(n,Q^2)$ in comparison to the exact analytical results. The
percentage error defined in (\ref{r4.3}) doesn't exceed 1\% in a case of the
flat input (\ref{r4.1}) and 2\% in a case of the more singular at small-$x$
input (\ref{r4.2}). The accuracy is better for lower $Q^2$, when the DGLAP
evolution is shorter. Using the results from Table 1, we expect the similar
precision for the truncated moments as well. Thus we assume that the Chebyshev 
method predictions are reliable with carefully estimated errors: 1\% for the 
parametrisation (\ref{r4.1}) and 2\% for (\ref{r4.2}). In Tables 2 and 3 we 
compare results for truncated at $x_0$ (0.01 and 0.1 respectively)  
moments, obtained from (\ref{r3.13}) (FMPR) with those, found 
within the Chebyshev approach (Cheb.). We set again two scales of $Q^2$: 10 
${\rm GeV}^2$ and 100 ${\rm GeV}^2$. 

\begin{table}[ht]
\begin{center}
\begin{tabular}{|c|c|c|c|c|}
\hline\hline
$q^{NS}(x,Q_0^2)$ & $Q^2$ & $n$ & $\bar{q}(Cheb.)$ &
$\bar{q}(FMPR)$ \\ \hline\hline
       &                   & 1 & $1.892\cdot 10^{-1}$ & $2.006\cdot 10^{-1}$ \\ 
\cline{3-5}
       &                   & 2 & $2.812\cdot 10^{-2}$ & $2.817\cdot 10^{-2}$ \\ 
\cline{3-5}
       &          100      & 3 & $7.499\cdot 10^{-3}$ & $7.491\cdot 10^{-3}$ \\ 
\cline{3-5}
       &                   & 4 & $2.740\cdot 10^{-3}$ & $2.732\cdot 10^{-3}$ \\ 
\cline{3-5}
$a_1 (1-x)^3$ &            & 5 & $1.212\cdot 10^{-3}$ & $1.204\cdot 10^{-3}$ \\ 
\cline{2-5}
       &                   & 1 & $1.951\cdot 10^{-1}$ & $2.015\cdot 10^{-1}$ \\ 
\cline{3-5}
       &                   & 2 & $3.289\cdot 10^{-2}$ & $3.293\cdot 10^{-2}$ \\ 
\cline{3-5}
       &          10       & 3 & $9.561\cdot 10^{-3}$ & $9.556\cdot 10^{-3}$ \\ 
\cline{3-5}
       &                   & 4 & $3.714\cdot 10^{-3}$ & $3.709\cdot 10^{-3}$ \\ 
\cline{3-5}
       &                   & 5 & $1.721\cdot 10^{-3}$ & $1.716\cdot 10^{-3}$ \\ 
\cline{1-5}
       &                   & 1 & $1.658\cdot 10^{-1}$ & $1.817\cdot 10^{-1}$ \\ 
\cline{3-5}
       &                   & 2 & $2.082\cdot 10^{-2}$ & $2.090\cdot 10^{-2}$ \\ 
\cline{3-5}
       &          100      & 3 & $5.250\cdot 10^{-3}$ & $5.245\cdot 10^{-3}$ \\ 
\cline{3-5}
       &                   & 4 & $1.907\cdot 10^{-3}$ & $1.902\cdot 10^{-3}$ \\ 
\cline{3-5}
$a_2 x^{-0.4}(1-x)^{2.5}$& & 5 & $8.550\cdot 10^{-4}$ & $8.502\cdot 10^{-4}$ \\ 
\cline{2-5}
       &                   & 1 & $1.732\cdot 10^{-1}$ & $1.826\cdot 10^{-1}$ \\ 
\cline{3-5}
       &                   & 2 & $2.437\cdot 10^{-2}$ & $2.443\cdot 10^{-2}$ \\ 
\cline{3-5}
       &          10       & 3 & $6.692\cdot 10^{-3}$ & $6.691\cdot 10^{-3}$ \\ 
\cline{3-5}
       &                   & 4 & $2.585\cdot 10^{-3}$ & $2.583\cdot 10^{-3}$ \\ 
\cline{3-5}
       &                   & 5 & $1.214\cdot 10^{-3}$ & $1.212\cdot 10^{-3}$ \\ 
\hline
\end{tabular}
\caption{Truncated at $x_0=0.01$ n-th moments $\bar{q}^{NS}(x_0,n,Q^2)$  within
FMPR and Chebyshev approaches for different $Q^2$ and input functions 
$q^{NS}(x,Q_0^2)$.}
\end{center}
\end{table}
\begin{table}[ht]
\begin{center}
\begin{tabular}{|c|c|c|c|c|}
\hline\hline
$q^{NS}(x,Q_0^2)$ & $Q^2$ & $n$ & $\bar{q}(Cheb.)$ &
$\bar{q}(FMPR)$ \\ \hline\hline
       &                   & 1 & $9.921\cdot 10^{-2}$ & $1.236\cdot 10^{-1}$ \\ 
\cline{3-5}
       &                   & 2 & $2.377\cdot 10^{-2}$ & $2.567\cdot 10^{-2}$ \\ 
\cline{3-5}
       &          100      & 3 & $7.229\cdot 10^{-3}$ & $7.368\cdot 10^{-3}$ \\ 
\cline{3-5}
       &                   & 4 & $2.721\cdot 10^{-3}$ & $2.724\cdot 10^{-3}$ \\ 
\cline{3-5}
 $a_1 (1-x)^3$ &           & 5 & $1.210\cdot 10^{-3}$ & $1.203\cdot 10^{-3}$ \\ 
\cline{2-5}
       &                   & 1 & $1.138\cdot 10^{-1}$ & $1.294\cdot 10^{-1}$ \\ 
\cline{3-5}
       &                   & 2 & $2.883\cdot 10^{-2}$ & $3.011\cdot 10^{-2}$ \\ 
\cline{3-5}
       &          10       & 3 & $9.303\cdot 10^{-3}$ & $9.401\cdot 10^{-3}$ \\ 
\cline{3-5}
       &                   & 4 & $3.695\cdot 10^{-3}$ & $3.699\cdot 10^{-3}$ \\ 
\cline{3-5}
       &                   & 5 & $1.719\cdot 10^{-3}$ & $1.715\cdot 10^{-3}$ \\ 
\cline{1-5}
       &                   & 1 & $7.112\cdot 10^{-2}$ & $9.086\cdot 10^{-2}$ \\ 
\cline{3-5}
       &                   & 2 & $1.664\cdot 10^{-2}$ & $1.816\cdot 10^{-2}$ \\ 
\cline{3-5}
       &          100      & 3 & $5.003\cdot 10^{-3}$ & $5.116\cdot 10^{-3}$ \\ 
\cline{3-5}
       &                   & 4 & $1.890\cdot 10^{-3}$ & $1.895\cdot 10^{-3}$ \\ 
\cline{3-5}
$a_2 x^{-0.4}(1-x)^{2.5}$& & 5 & $8.536\cdot 10^{-4}$ & $8.497\cdot 10^{-4}$ \\ 
\cline{2-5}
       &                   & 1 & $8.237\cdot 10^{-2}$ & $9.519\cdot 10^{-2}$ \\ 
\cline{3-5}
       &                   & 2 & $2.026\cdot 10^{-2}$ & $2.131\cdot 10^{-2}$ \\ 
\cline{3-5}
       &          10       & 3 & $6.446\cdot 10^{-3}$ & $6.528\cdot 10^{-3}$ \\ 
\cline{3-5}
       &                   & 4 & $2.568\cdot 10^{-3}$ & $2.572\cdot 10^{-3}$ \\ 
\cline{3-5}
       &                   & 5 & $1.213\cdot 10^{-3}$ & $1.211\cdot 10^{-3}$ \\ 
\hline
\end{tabular}
\caption{Truncated at $x_0=0.1$ n-th moments $\bar{q}^{NS}(x_0,n,Q^2)$  within
FMPR and Chebyshev approaches for different $Q^2$ and input functions 
$q^{NS}(x,Q_0^2)$.}
\end{center}
\end{table}
\begin{figure}[ht]
\begin{center}
\includegraphics[width=90mm]{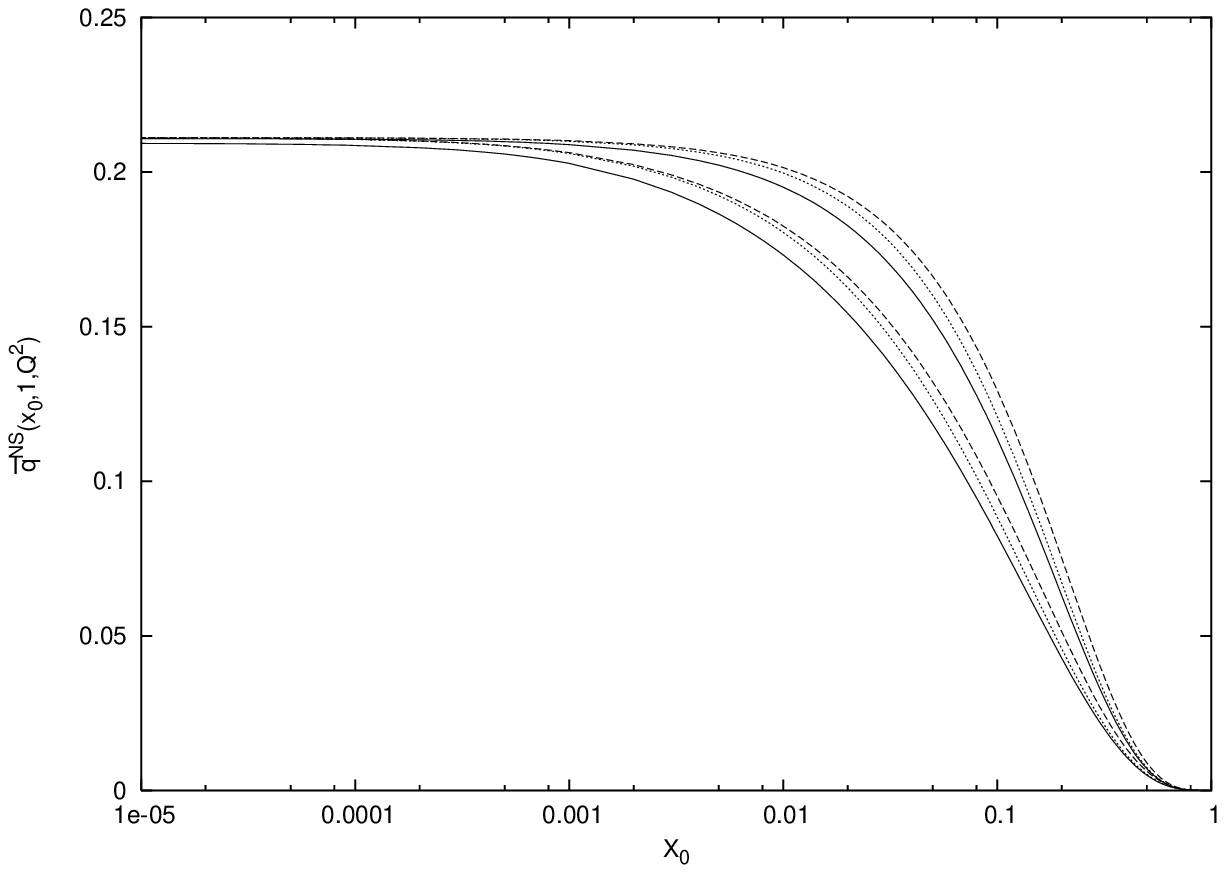}
\caption{First truncated moment: $\bar{q}^{NS}(x_0,1,Q^2)Cheb.$ (solid),
$\bar{q}^{NS}(x_0,1,Q^2)FMPR$ (dashed $M=4$, dotted $M=20$) 
for different inputs: (\ref{r4.1}) - upper lines and 
(\ref{r4.2}) - lower lines. $Q^2=10 {\rm GeV}^2$.}
\end{center}
\end{figure}
\begin{figure}[ht]
\begin{center}
\includegraphics[width=90mm]{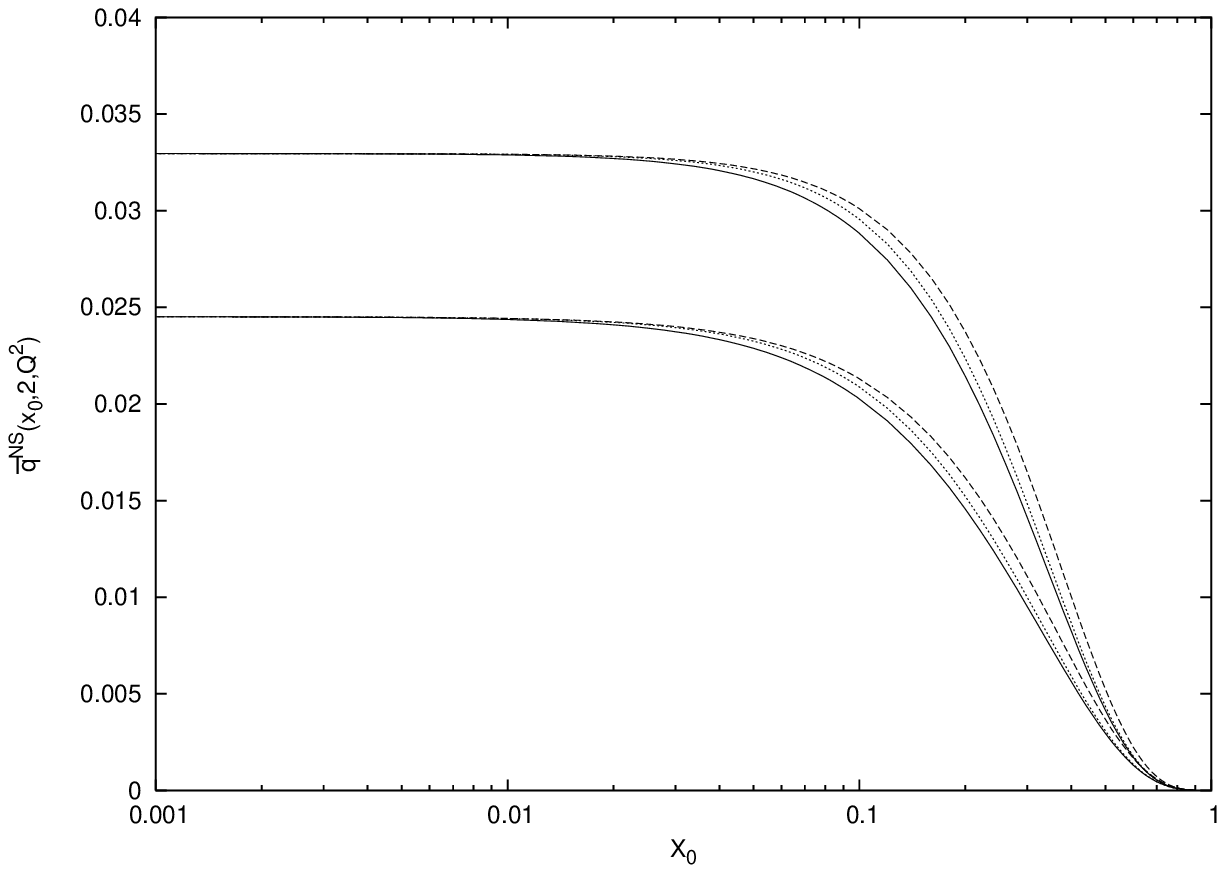}
\caption{Second truncated moment: $\bar{q}^{NS}(x_0,2,Q^2)Cheb.$ (solid),
$\bar{q}^{NS}(x_0,2,Q^2)FMPR$ (dashed $M=4$, dotted $M=20$) 
for different inputs: (\ref{r4.1}) - upper lines and 
(\ref{r4.2}) - lower lines. $Q^2=10 {\rm GeV}^2$.}
\end{center}
\end{figure}
\begin{figure}[ht]
\begin{center}
\includegraphics[width=90mm]{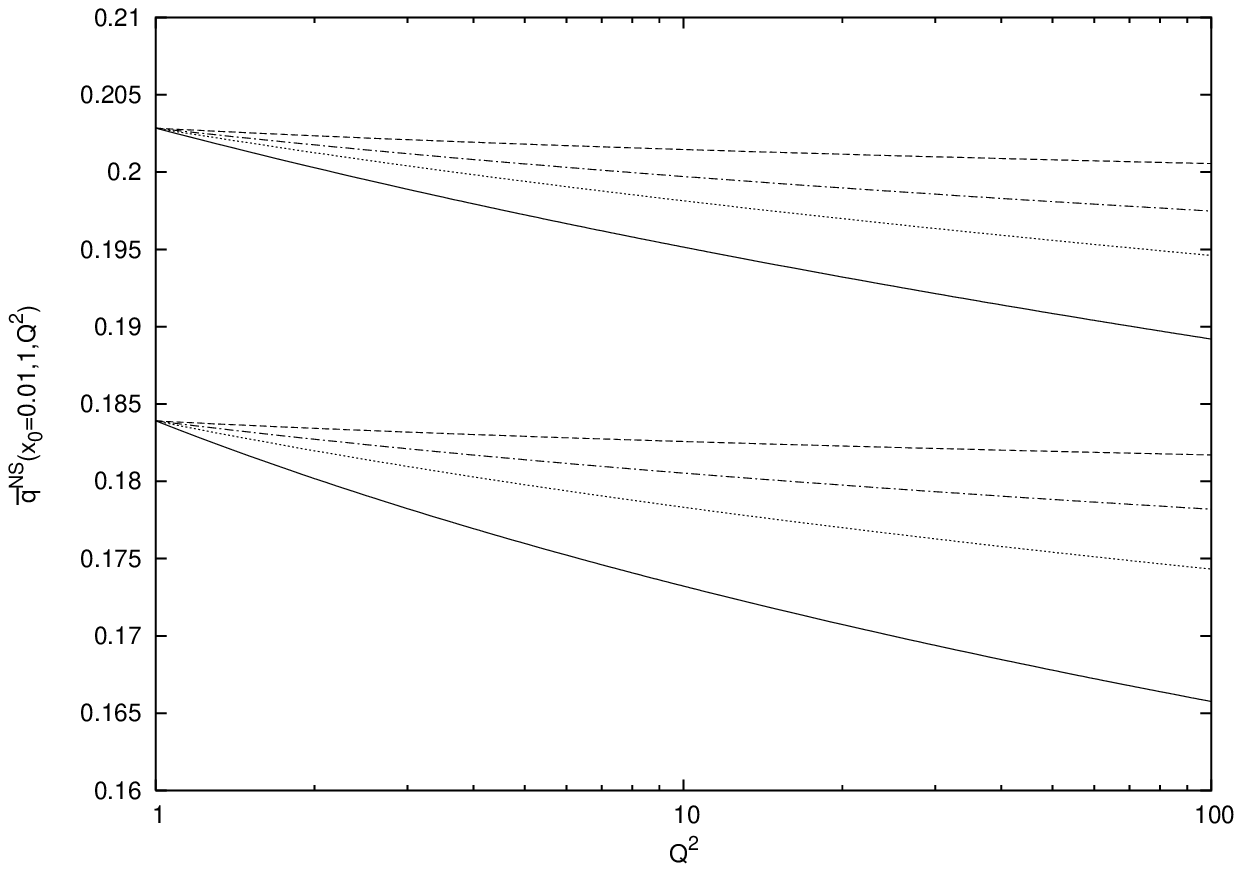}
\caption{First truncated at $x_0=0.01$ moment: $\bar{q}^{NS}(x_0,1,Q^2)Cheb.$ 
(solid), $\bar{q}^{NS}(x_0,1,Q^2)FMPR$ (dashed $M=4$, dashed-dotted $M=20$, 
dotted $M=60$). The upper lines correspond to the input 
parametrisation (\ref{r4.1}), the lower ones to (\ref{r4.2}).}
\end{center}
\end{figure}
\begin{figure}[ht]
\begin{center}
\includegraphics[width=90mm]{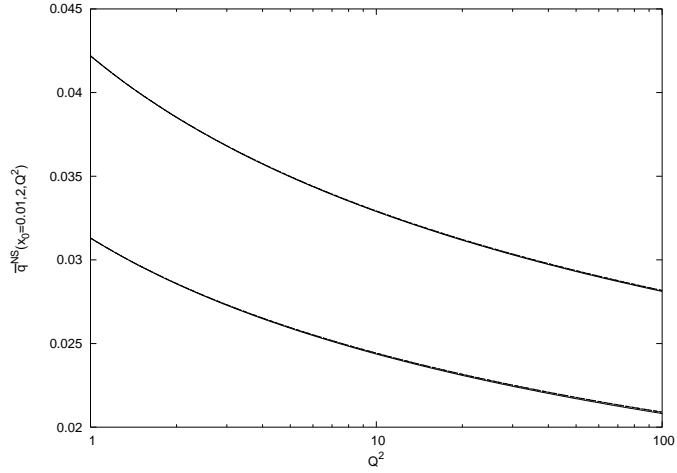}
\caption{Second truncated at $x_0=0.01$ moment: $\bar{q}^{NS}(x_0,2,Q^2)Cheb.$ 
(solid), $\bar{q}^{NS}(x_0,2,Q^2)FMPR$ (covered with Cheb. for different
$M\geq 4$). The upper line corresponds to the input parametrisation (\ref{r4.1}), 
the lower one to (\ref{r4.2}).}
\end{center}
\end{figure}
\begin{figure}[ht]
\begin{center}
\includegraphics[width=90mm]{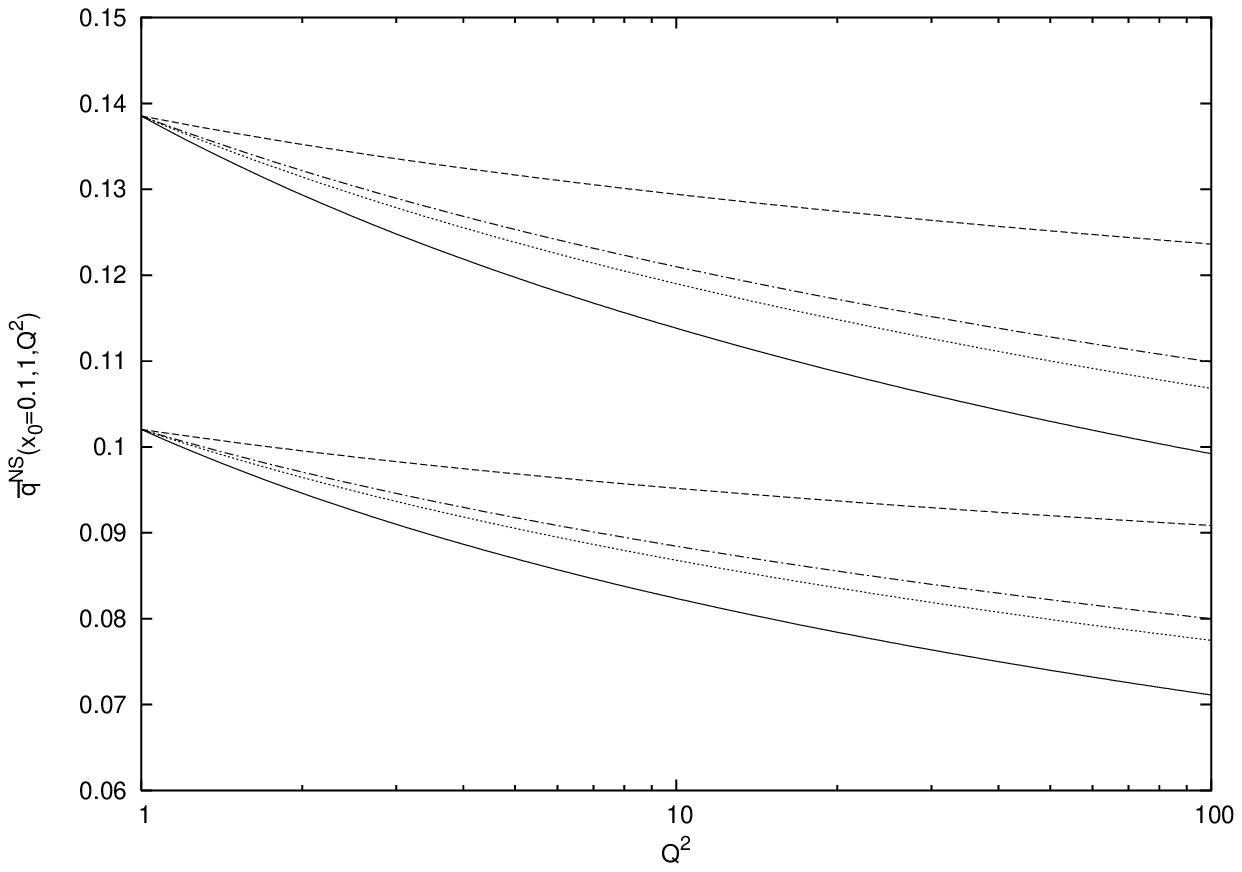}
\caption{First truncated at $x_0=0.1$ moment: $\bar{q}^{NS}(x_0,1,Q^2)Cheb.$ 
(solid), $\bar{q}^{NS}(x_0,1,Q^2)FMPR$ (dashed $M=4$, dashed-dotted $M=20$, 
dotted $M=60$). The upper lines correspond to the input 
parametrisation (\ref{r4.1}), the lower ones to (\ref{r4.2}).}
\end{center}
\end{figure}
\begin{figure}[ht]
\begin{center}
\includegraphics[width=90mm]{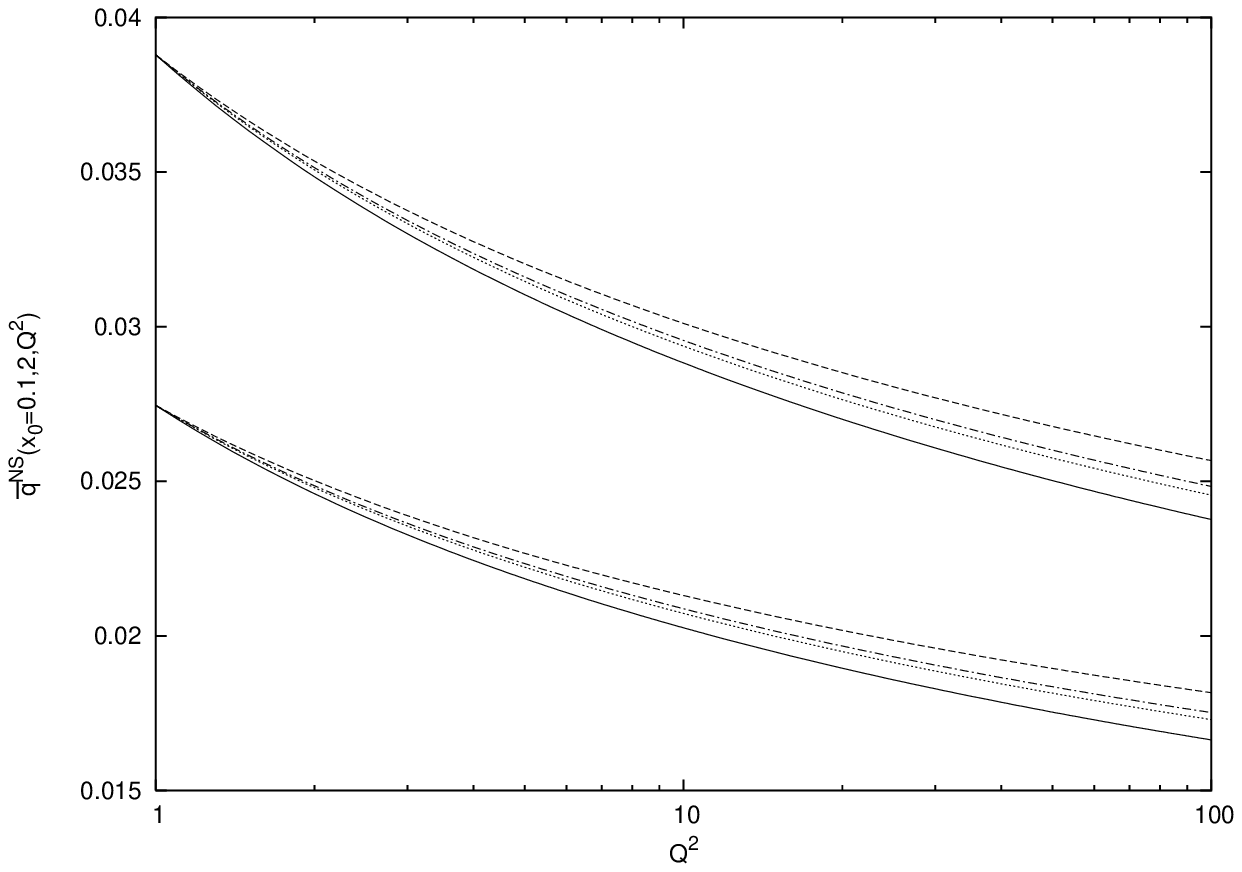}
\caption{Second truncated at $x_0=0.1$ moment: $\bar{q}^{NS}(x_0,2,Q^2)Cheb.$ 
(solid), $\bar{q}^{NS}(x_0,2,Q^2)FMPR$ (dashed $M=4$, dashed-dotted $M=20$, 
dotted $M=60$). The upper lines correspond to the input 
parametrisation (\ref{r4.1}), the lower ones to (\ref{r4.2}).}
\end{center}
\end{figure}
\noindent Notice a quite satisfactory agreement of the both presented
methods even for a very small value of $M$ (4).
The accuracy of the determination of higher moments is better despite the
fact, that less terms $(M-n)$ are included. The accuracy of the truncated 
moments method depends on the convergence of the expansion of $G_n(x_0/y)$, 
which is the truncated counterpart of the anomalous dimension $\gamma_{qq}(n)$. 
Because $G_n(x_0/y)$ is expanded in powers of $y$ around $y=1$, 
the small-$y$ region ($y\sim x_0$) in the integral of the evolution equation 
(\ref{r3.4}) is badly reproduced. Therefore the convergence is better for 
higher moments, which have a smaller contribution from the low-$y$ region. 
Lower moments are more sensitive to the lower limit of the integration $x_0$ in 
(\ref{r3.4}). From the other side, for sufficiently small $x_0$, factors 
$x_0^i$ in the coefficients $C_{pn}^M(x_0)$ (\ref{r3.10}) make the convergence 
of $G_n(x_0/y)$ better. Hence the difference 
between $\bar{q}^{NS}(x_0,n,Q^2)FMPR$ and $\bar{q}^{NS}(x_0,n,Q^2)Cheb.$ is 
larger for $x_0=0.1$ than for $x_0=0.01$. Furthermore, as $x_0\rightarrow 1$,
the accordance of $\bar{q}^{NS}(x_0,n,Q^2)FMPR$ and 
$\bar{q}^{NS}(x_0,n,Q^2)Cheb.$ becomes again better because of the vanishing
structure functions in this limit. Comparisons of $\bar{q}^{NS}(x_0,n,Q^2)FMPR$ 
with $\bar{q}^{NS}(x_0,n,Q^2)Cheb.$ as a function of $x_0$ for first ($n=1$)
and second ($n=2$) moments are shown in Figs.1,2. In Figs.3-6 we present the
$Q^2$ dependence of $\bar{q}^{NS}(x_0,n,Q^2)FMPR$ and 
$\bar{q}^{NS}(x_0,n,Q^2)Cheb.$ at fixed $x_0=0.01, 0.1$ and for $n=1$,
$n=2$ respectively. The plots are given for different $M$ and both
parametrisations (\ref{r4.1}),(\ref{r4.2}). The agreement of the truncated
moment method with the Chebyshev approach is perfect for $n=2$ at
$x_0\leq 0.01$, independently on the inputs, $Q^2$ and even value of $M$. The
other results are also very satisfactory. The relative difference between 
$\bar{q}^{NS}(x_0,n,Q^2)FMPR$ and $\bar{q}^{NS}(x_0,n,Q^2)Cheb.$ doesn't
exceed 5\% for $n\geq 2$ and not too large $x_0$ ($x_0\leq 0.1$), already at
$M=4$. This difference for the first moment also decreases down to a few \%
for $M=30$ and $x_0=0.1$ (for smaller $x_0$ the accuracy is much better).
The error function 
\begin{equation}\label{r4.3a}
R_n^M(x_0,Q^2) =
\frac{\mid\bar{q}^{NS}(x_0,n,Q^2)FMPR-\bar{q}^{NS}(x_0,n,Q^2)Cheb.\mid}
{\bar{q}^{NS}(x_0,n,Q^2)Cheb.}\cdot 100\%
\end{equation}
grows very slowly with $Q^2$ (see Figs.3,5,6). In Tables 4 and 5 we show
the error function $R_n^M(x_0,Q^2)$ for different $M$, $Q^2=10 {\rm GeV}^2$
and two values of $x_0: 0.01, 0.1$ respectively.
\begin{table}[ht]
\begin{center}
\begin{tabular}{|c|c|c|c|c|}
\hline\hline
$x_0=0.01$ & \multicolumn{2}{c|}{$a_1 (1-x)^{3}$} & \multicolumn{2}{c|}
{$a_2 x^{-0.4}(1-x)^{2.5}$} \\ \hline\hline
$M$ & $~R_1~$ & $R_2$ & $~~~R_1~~~$ & $R_2$ \\ \hline
 4 & 4     & $\ll 1$ & 6 & $\ll 1$ \\ \hline
10 & 3     & $\ll 1$ & 5 & $\ll 1$ \\ \hline
20 & 3     & $\ll 1$ & 5 & $\ll 1$ \\ \hline
30 & 2     & $\ll 1$ & 4 & $\ll 1$ \\ \hline
60 & 2     & $\ll 1$ & 3 & $\ll 1$ \\ \hline
\end{tabular}
\caption{The percentage error function $R_n\equiv R_n^M(x_0,Q^2)$ defined in 
(\ref{r4.3a}), for $x_0=0.01$ and different input functions $q^{NS}(x,Q_0^2)$. 
$Q^2=10 {\rm GeV}^2$, the values of $n$ and $M$ shown.}
\end{center}
\end{table}
\begin{table}[ht]
\begin{center}
\begin{tabular}{|c|c|c|c|c|}
\hline\hline
$x_0=0.1$ & \multicolumn{2}{c|}{$a_1 (1-x)^{3}$} & \multicolumn{2}{c|}
{$a_2 x^{-0.4}(1-x)^{2.5}$} \\ \hline\hline
$M$ & $~R_1~$ & $R_2$ & $~~~R_1~~~$ & $R_2$ \\ \hline
 4 & 13   & 5 & 16 & 5 \\ \hline
10 & 10   & 4 & 11 & 4 \\ \hline
20 &  6   & 3 &  7 & 3 \\ \hline
30 &  4   & 2 &  5 & 2 \\ \hline
\end{tabular}
\caption{The percentage error function $R_n\equiv R_n^M(x_0,Q^2)$ defined in 
(\ref{r4.3a}), for $x_0=0.1$ and different input functions $q^{NS}(x,Q_0^2)$. 
$Q^2=10 {\rm GeV}^2$, the values of $n$ and $M$ shown.}
\end{center}
\end{table}
Note that with increasing $M$ the accuracy of the truncated moments method
systematically though slowly increases. This improvement of the accuracy
breaks however for larger $M$ ($M\simeq 70$ at $x_0=0.01$ and $M\simeq 40$ at 
$x_0=0.1$ ) because of increasing numerical errors. All presented above results
concern the running coupling $\alpha_s(Q^2)$. We have found also, that for
the constant $\alpha_s$ the error function $R$ (\ref{r4.3a}) grows
approximately proportionally to the strength of $\alpha_s$:
\begin{equation}\label{r4.3b}
\frac{R(\alpha_{s1})}{R(\alpha_{s2})} \sim \frac{\alpha_{s1}}{\alpha_{s2}}.
\end{equation}
Summarising, the LO DGLAP evolution of any truncated at $x_0\leq 0.1$ moment
of the parton distribution can be reproduced with the satisfactory accuracy, 
where the relative error $\leq 5 \%$.

\section{Summary and conclusions.}

Analysis of the QCD $Q^2$ evolution equations for truncated moments of
parton distributions is very interesting both from the theoretical and
experimental point of view. The truncated moments technique is complementary
to the existing methods for solving the evolution equations, based on the
full moments or $x$-space approaches. Apart from it refers directly to the
physical values - moments (rather than to the parton distributions), what
enables to use a wide range of deep-inelastic scattering data in terms of
smaller number of parameters. In this way, no assumptions on the shape of
parton distributions are needed. Dealing with truncated at $x_0$ Mellin 
moments: $\int_{x_0}^{1} dx x^{n-1} f(x,Q^2)$ one can also avoid 
uncertainty from the unmeasurable very small $x\rightarrow 0$ region. 

In this paper we have compared the solutions of LO DGLAP $Q^2$ evolution 
equations written for the truncated Mellin moments of the structure functions 
with those, obtained by using the Chebyshev-polynomial technique. 
In both these approaches we have calculated numerically and
semi-analytically the truncated moments $\int_{x_0}^{1} dx x^{n-1} 
F(x,Q^2)$. As a test structure function $F(x,Q^2)$ we have taken two different 
spin-like nonsinglet parton distributions. The computations have been
performed for a wide range of $x_0$ ($10^{-5}\leq x_0\leq 0.9$) and $Q^2$ 
($1\leq Q^2\leq 100 {\rm GeV}^2$). Treating the Chebyshev results
as exact, we have found that the truncated moments method is very promising,
for any moment, together with the first one. The precision of the truncated 
moments approach is perfect for higher moments ($n\geq 2$) and not too large 
the truncation point $x_0$ ($x_0\leq 0.1$), even for small $M=4$. Larger
values of $M$ (e.g. $M=30$) enables to obtain a quite satisfactory accuracy
(the relative error $\leq 5\%$) also for the first truncated moment. The 
original truncated moments technique \cite{b5} has been developed in \cite{b5a}, 
what could improve the numerical efficiency. This technique can be a valuable 
tool e.g. in determination of the contribution to the moments of the gluon 
distribution from the experimentally accessible region. We think that the 
method of truncated moments can be useful in further theoretical and 
experimental QCD investigations.

\appendix
\section{Chebyshev polynomial expansion within LO DGLAP evolution equations.}
 
In order to solve the integro-differential evolution equation 
\begin{equation}\label{rA.7}
\frac{\partial q^{NS}(x,t)}{\partial t}=\frac{\alpha_s(t)}{2\pi}
\int\limits_x^1\frac{dz}{z} P_{qq}\left(\frac{x}{z}\right) q^{NS}(z,t)
\end{equation}
one has to expand functions $q^{NS}(x,t)$ into the series of the Chebyshev
polynomials:
\begin{equation}\label{rA.1}
q^{NS}(x,t) \rightarrow q^{NS}(x',t) = \frac{2}{N}\sum\limits_{i=0}^{N-1}
\sum\limits_{k=0}^{N-1}\upsilon_i \:q^{NS}(x_k,t)T_i(\tau_k)T_i(x'),
\end{equation}
where
\begin{equation}\label{rA.2}
\upsilon_i = \cases{0.5 & for~~ $i=0$ \cr 1 & for~~ $i\geq 1$ \cr},
\end{equation}
\begin{equation}\label{rA.3}
x' = \frac{2\ln x}{\ln x_{min}} - 1.
\end{equation}
$T_i(x)$ is the Chebyshev polynomial, defined as \cite{b15}:
\begin{equation}\label{rA.4}
T_i(x) = \cos (i\arccos (x))
\end{equation}
and $\tau_k$ are nodes (zeros) of the $T_n$:
\begin{equation}\label{rA.5}
\tau_k = \cos\frac{2k+1}{2n}\pi,~~~~~k=0,1,2,...,n-1,
\end{equation}
\begin{equation}\label{rA.6}
x_k = x_{min}^{0.5(\tau_k + 1)}.
\end{equation}
$x_{min}$ in (\ref{rA.3}) and (\ref{rA.6}) is the smallest value of Bjorken
$x$, involved in the analysis. In our computations $x_{min}=10^{-6}$.
Transformation (\ref{rA.3}) converts the physical $x$-region: [$x_{min};1$]
into the $x'\in [-1;1]$ one, suitable for the Chebyshev approximation.
Integration over $z$ in the evolution equation (\ref{rA.7}) with $q^{NS}(x,t)$ 
expanded according to (\ref{rA.1}) leads to the system of linear differential 
equations:
\begin{equation}\label{rA.8}
\frac{dq^{NS}(x_i,t)}{dt} = \sum\limits_{j=0}^{N-1} H_{ij} q^{NS}(x_j,t).
\end{equation}
This system can be solved by using the standard Runge-Kutta method with
initial conditions given by the input parametrisation $q^{NS}(x_j,t_0)$. $N$
in the polynomial expansion (\ref{rA.1}) is equal to 20.

\section{Analytical solution of the system of DGLAP evolution equations for
truncated moments $\bar{q}^{NS}(x_0,n,t)$.}

The closed system of $M+1$ DGLAP evolution equations for truncated moments 
$\bar{q}^{NS}(x_0,n,t)$ (\ref{r3.11}) can be rewritten in the form:
\begin{displaymath}
\frac{d\bar{q}^{NS}(x_0,N_0,t)}{dt}=\frac{\alpha_s(t)}{2\pi}
[D_{N_0,N_0}^{(M)}(x_0) \bar{q}^{NS}(x_0,N_0,t)
\end{displaymath}
\begin{displaymath}
+ D_{N_0,N_0+1}^{(M)}(x_0) \bar{q}^{NS}(x_0,N_0+1,t) +
...+ D_{N_0,N_0+M}^{(M)}(x_0) \bar{q}^{NS}(x_0,N_0+M,t)]
\end{displaymath}
\begin{displaymath}
\frac{d\bar{q}^{NS}(x_0,N_0+1,t)}{dt}=\frac{\alpha_s(t)}{2\pi}
[D_{N_0+1,N_0+1}^{(M-1)}(x_0) \bar{q}^{NS}(x_0,N_0+1,t)
\end{displaymath}
\begin{displaymath}
+ D_{N_0+1,N_0+2}^{(M-1)}(x_0) \bar{q}^{NS}(x_0,N_0+2,t) +
...+ D_{N_0+1,N_0+M}^{(M-1)}(x_0) \bar{q}^{NS}(x_0,N_0+M,t)]
\end{displaymath}
\begin{displaymath}
...
\end{displaymath}
\begin{equation}\label{rB.1}
\frac{d\bar{q}^{NS}(x_0,N_0+M,t)}{dt}=\frac{\alpha_s(t)}{2\pi}
D_{N_0+M,N_0+M}^{(0)}(x_0) \bar{q}^{NS}(x_0,N_0+M,t).
\end{equation}
$N_0$ denotes the lowest moment in calculations and the matrix elements 
$D_{ij}^{(k)}(x_0)$ are related to the $C_{ij}^{(k)}(x_0)$ (\ref{r3.10}) via
\begin{equation}\label{rB.2}
D_{ij}^{(k)}(x_0) = \cases{ C_{j-i,i}^{(k)}(x_0) & $j\geq i$ \cr 0 & $j<i$
\cr}.
\end{equation}
$D$ is a triangular matrix and therefore (\ref{rB.1}) can be solved 
analytically using the diagonalising matrix A:
\begin{equation}\label{rB.3}
A_{ij}(x_0) =
\frac{D_{ij}(x_0)-\sum\limits_{k=i+1}^{j-1}D_{kj}(x_0)A_{ik}}{D_{jj}(x_0)-
D_{ii}(x_0)}.
\end{equation}
In this way one obtains the recurrence solutions (\ref{r3.12}),(\ref{r3.13}).


\begin{thebibliography}{99}

\bibitem{b1} V.N.Gribov, L.N.Lipatov, {\it Sov. J. Nucl. Phys.}
{\bf 15}, 438 and 675 (1972); Yu.L.Dokshitzer, {\it Sov. Phys. JETP}
{\bf 46}, 641 (1977); G.Altarelli, G.Parisi, {\it Nucl. Phys.} {\bf B126}, 298
(1977).
\bibitem{b4} J.Kwieci\'nski, M.Maul, {\it Phys. Rev.} {\bf D67}, 034014 (2003).
\bibitem{b5} S.Forte, L.Magnea, {\it Phys. Lett.} {\bf B448}, 295 (1999); 
S.Forte, L.Magnea, A.Piccione, G.Ridolfi, {\it Nucl. Phys.} {\bf B594}, 46 
(2001).
\bibitem{b5a} A.Piccione, {\it Phys. Lett.} {\bf B518}, 207 (2001).
(2001).
\bibitem{b6} D.Kotlorz, A.Kotlorz, {\it Acta Phys. Pol.} {\bf B35}, 705 (2004).
\bibitem{b7} P.Santorelli, E.Scrimieri, {\it JHEP Conf.Proc.corfu} 98/023, 
{\tt hep-ph/9909289}.
\bibitem{b8} S.Kumano, T.-H.Nagai, {\it J.Comput. Phys.} {\bf 201}, 651 (2004), 
{\tt hep-ph/0405160}.
\bibitem{b12} J.Kwieci\'nski, D.Str\'ozik-Kotlorz, {\it Z. Phys.} {\bf C48}, 
315 (1990);
J.Kwieci\'nski, B.Ziaja, {\it Phys. Rev.} {\bf D60}, 054004 
(1999); B.Bade\l{}ek, J.Kwieci\'nski, {\it Phys. Lett.} {\bf B418}, 229 (1998).
\bibitem{b14} B.I.Ermolaev, M.Greco, S.I.Troyan, {\it Nucl. Phys.} {\bf B571}, 
137 (2000); {\tt hep-ph/0106317}; {\it Nucl. Phys.} {\bf B594}, 71 (2001); 
{\it Phys. Lett.} {\bf B522}, 57 (2001); {\it Phys. Lett.} {\bf B579}, 321 
(2004); {\tt hep-ph/0404267}.
\bibitem{b15} S.E.El-gendi, {\it Chebyshev solution of differential,
integral and integro-differential equations, Comput. J.} {\bf 12}, 282-287
(1969).

\end{thebibliography}
\end{document}